\documentclass[a4paper,11pt,toc=bibliography]{article}
\usepackage{pos}
\usepackage{siunitx}
\usepackage{todonotes}
\usepackage{url}
\usepackage{hyperref}[hidelinks]
\usepackage{doi}
\usepackage{nicefrac}
\usepackage{caption}
\usepackage{subcaption}
\usepackage{graphicx}
\usepackage{nicefrac}

\usepackage{csquotes}

\title{Analysis of the Supernova Remnant IC 443 using H.E.S.S. Data}

\author*[a]{Alison M. W. Mitchell}
\author[a]{Lukas Großpietsch}
\author[a]{Tina Wach}

\onbehalf{for the H.E.S.S. Collaboration}

\affiliation[a]{Erlangen Centre for Astroparticle Physics, Friedrich-Alexander-Universität Erlangen-Nürnberg,\\
Nikolaus-Fiebiger-Straße 2, D 91058 Erlangen, Germany}

\emailAdd{alison.mw.mitchell@fau.de}

\abstract{IC 443 is a well-known supernova remnant that stands out due to its interaction with a dense molecular cloud, creating a complex environment where shocks can efficiently accelerate particles to high energies. This makes it a key target for investigating the mechanisms of cosmic-ray acceleration and gamma-ray production, particularly in the context of supernova remnants as potential sources of PeV cosmic rays. This work presents a first analysis of the region as observed by H.E.S.S.. We detect extended very-high-energy gamma-ray emission from IC 443, consistent with previous observations by VERITAS and MAGIC. A multi-wavelength comparison incorporating data from Fermi-LAT, MAGIC, and VERITAS strongly supports a hadronic origin of the observed emission, and highlights the presence of relativistic protons interacting with the surrounding molecular cloud. These findings reinforce the role of IC 443 as a key laboratory for studying supernova remnants as cosmic-ray accelerators and their interaction with their surrounding mediums.}

\ConferenceLogo{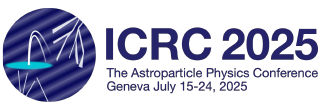}

\FullConference{39th International Cosmic Ray Conference (ICRC2025)\\
 15–24 July 2025\\
Geneva, Switzerland\\}

\begin{document}
\maketitle

\section{Introduction}

Supernova remnants (SNRs) are widely regarded as potential sources of Galactic cosmic rays, capable of accelerating particles to relativistic energies through diffusive shock acceleration. The interaction of these high-energy particles with ambient gas and radiation fields produces $\gamma$-ray emission across a wide energy range, providing a critical probe into the nature and efficiency of particle acceleration processes.

IC\,443 (G189.1+3.0) is a well-studied middle-aged SNR located at a distance of approximately 1.5 kpc in the Galactic anticenter region \cite{fesen_nature_1984}. The age of the remnant is still discussed, but is predominantly reported in the range of $20-\SI{30}{kyr}$ \cite{fesen_nature_1984,torres_supernova_2003,olbert_bow_2001}



Previous observations from instruments such as Fermi-LAT and VERITAS have revealed both GeV and TeV gamma-ray emission from this source. IC\,443 has been first observed in gamma rays by EGRET \cite{esposito_egret_1996} and later by Fermi-LAT \cite{Fermi_2010,Fermi_2013}, MAGIC \cite{MAGIC_2007}, and VERITAS \cite{VERITAS_2009}.
The source region derived by MAGIC coincides with a large molecular cloud and is shifted from the center positions derived by Fermi-LAT and VERITAS. However, the mostly overlapping regions derived by VERITAS and Fermi-LAT are more spatially extended, incorporating most molecular clouds and the inner shell of the supernova remnant, as well as the pulsar and the pulsar wind nebula. The analysis of the Fermi-LAT observations of the source report a broad peak of the emission between a few $\SI{100}{MeV}$ and $\SI{5}{GeV}$ which is consistent with an emission caused by neutral pion decay emitting $\gamma$-rays \cite{Fermi_2010, Fermi_2013, MAGIC_2007, VERITAS_2009}. This detection makes the remnant one of the first objects for observationally proving the interaction between a core-collapse SNR and a dense molecular cloud, which in turn makes it an excellent candidate for studying hadronic acceleration mechanisms.

This work presents an analysis of data from the High Energy Stereoscopic System (H.E.S.S.), an Imaging Atmospheric Cherenkov Telescope (IACT) array located in the Khomas Highlands in Namibia (23$^\circ$16'16.79''\,S, 16$^\circ$30'0''\,E). The results are compared with previous observations from other $\gamma$-ray instruments, and the corresponding spectrum of the parent proton population is derived.

\section{H.E.S.S. Data Analysis and Results}
The data used in this work were quality selected such that only observations for which all 4 telescopes were operational were included because all data analyzed in this work were taken before completion of the fifth H.E.S.S. IACT. Additionally, the quality criteria detailed in \cite{the_hess_collaboration_observations_2006} were applied, to ensure a good system response as well as good atmospheric conditions for every observation. This selection yields a dataset with a total deadtime corrected observation time of 11.4 hours between December 2004 and December 2005. The data include observations of IC\,443 in a zenith range of $41.1^\circ$ to $54.7^\circ$ with an average of $47.9^\circ$. The energy threshold is derived as the energy above the peak of the background spectrum with an energy bias below $10\%$. This yields an energy threshold of $\SI{0.56}{TeV}$. The analysis presented in this work is performed using version 1.2 of the open source Python package \texttt{gammapy} \cite{donath_gammapy_2023, gammapy_zenodo}.

\begin{figure*}
\begin{minipage}[b]{.49\textwidth}
\includegraphics[width=0.99\textwidth]{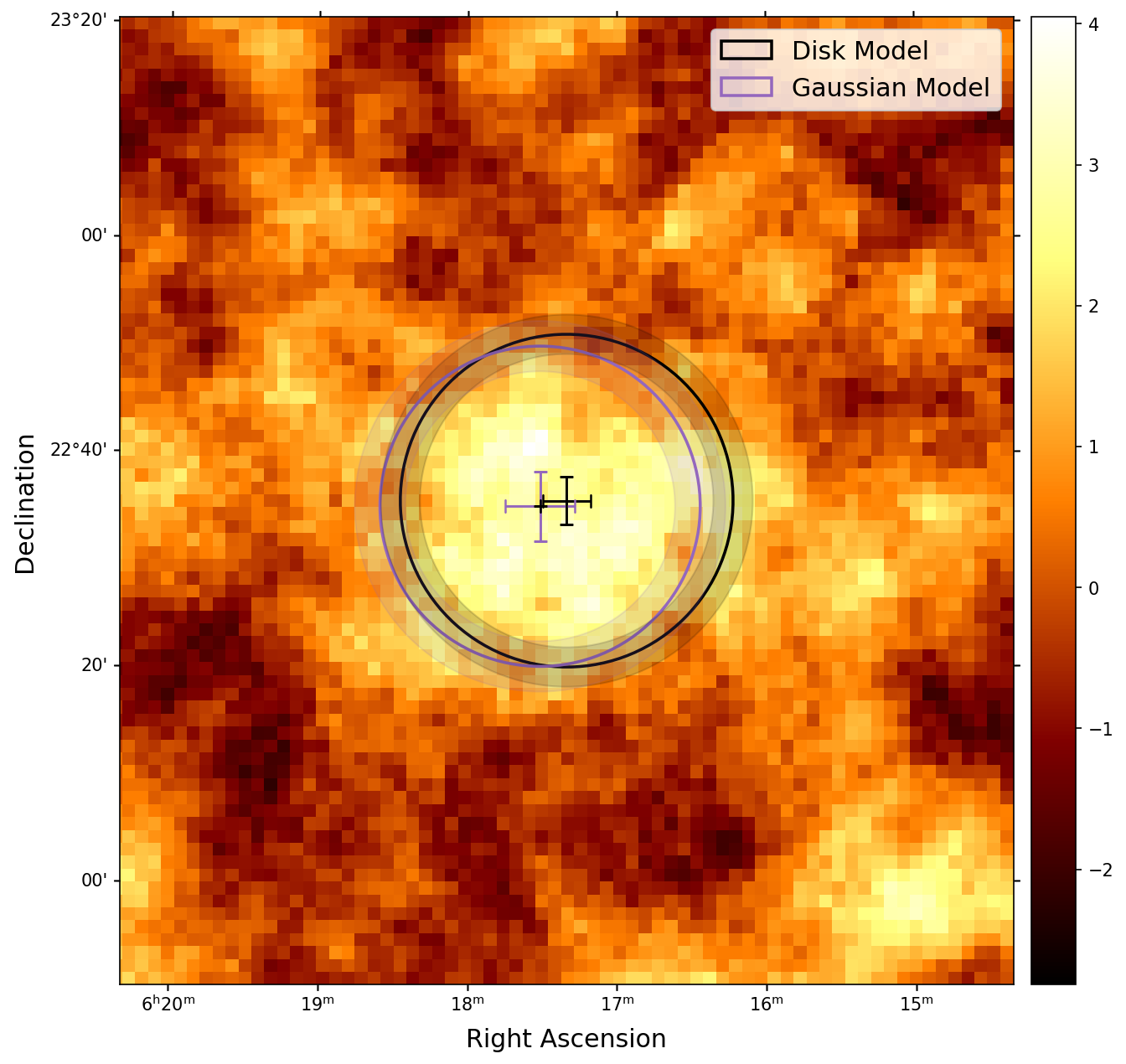}
\end{minipage}\qquad
\begin{minipage}[b]{.49\textwidth}
\includegraphics[width=0.92\textwidth]{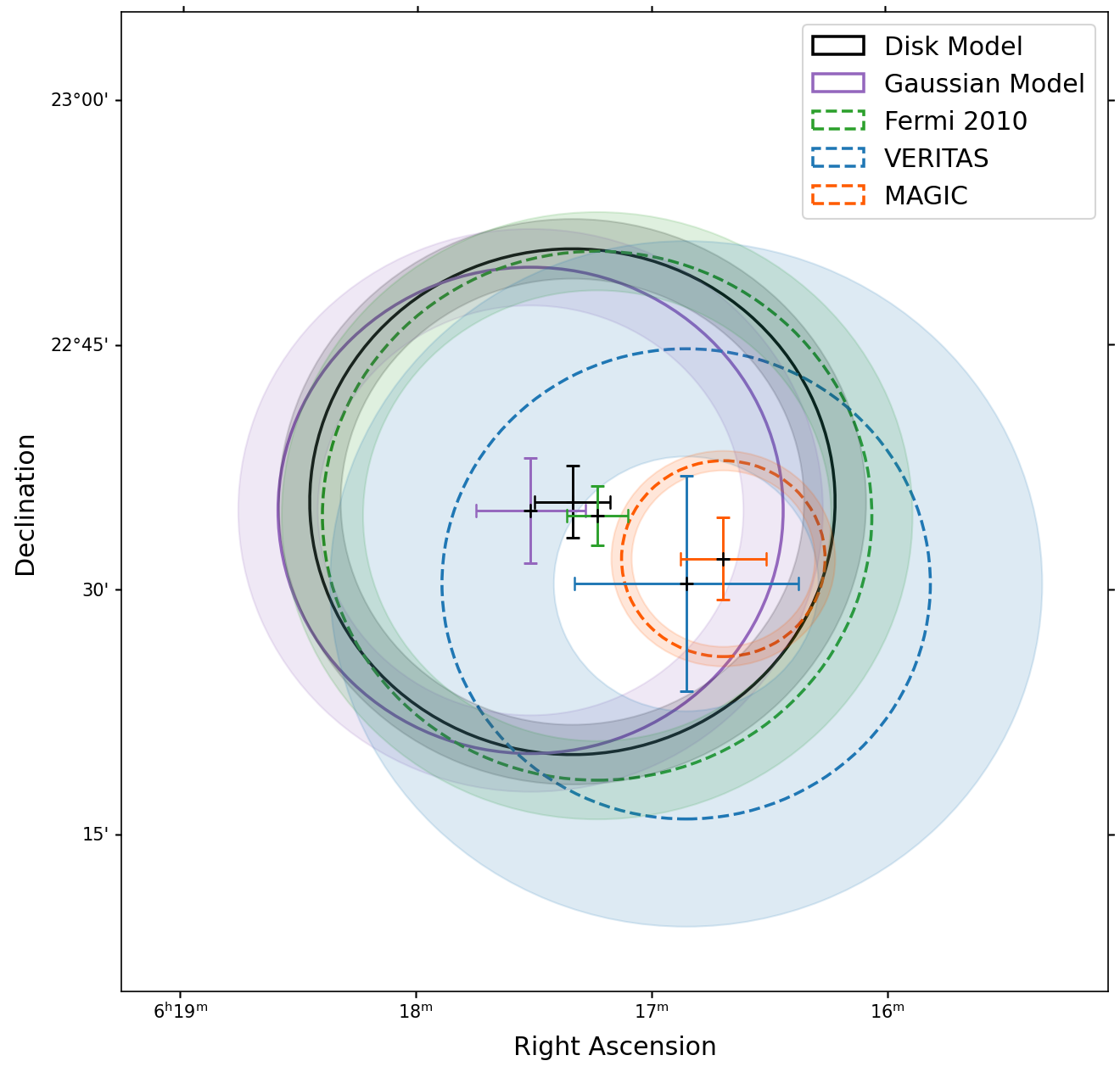}
\end{minipage}
\caption{Left: Li\&Ma significance map of the region around IC\,443, computed with a correlation radius of $0.16\deg$. The best-fit morphology of the Gaussian and Disk models are shown by the purple and black circle. Right: The best-fit morphology of both models compared to the emission observed by other instruments. The uncertainties, depicted as error bars for the position and shaded regions for the extension, include both statistical and systematic uncertainties (H.E.S.S. systematics from \cite{collaboration_resolving_2019}).}
\label{fig:morphology}
\end{figure*}

Gamma-hadron separation was performed using a method described in \citep{gamma-hadron} and reconstructed using the Image Pixel-wise fit for Atmospheric Cherenkov Telescopes algorithm \texttt{ImPACT} (for more information, see \citep{impact}). This modern reconstruction framework offers significantly improved angular resolution and background rejection compared to the analysis methods available at the time the data were originally taken. As a result, archival observations from 2004 and 2005 can now be revisited with enhanced sensitivity, allowing us to uncover previously inaccessible signals or faint excesses. The analysis was cross-checked with an independent reconstruction chain \cite{model,de_naurois_high_2009}. The background has been estimated using a background model template derived from a large set of archival observations \cite{mohrmann_validation_2019}.

A Li\&Ma \cite{li_analysis_1983} significance map of the region centered on IC\,443 can be seen in the left panel of Figure \ref{fig:morphology}, showing extended $\gamma$-ray emission around the nominal SNR position.
\begin{table}
\caption{Best-fit parameters obtained for the analysis of the data above the energy threshold, using a Gaussian and a Disk model with a power law as spectral model. For the disk model, $\sigma$ gives the radius of the fitted disk $r_0$. The quoted uncertainties do not include systematics.} 
\label{tab:fitparamters} 
\centering                          
\begin{tabular}{c | c  c}       
\hline\hline  
\noalign{\smallskip}
 & Disk Model & Gaussian Model  \\    
\noalign{\smallskip}
\hline 
\noalign{\smallskip}
   $\Gamma$  & $3.5 \pm  0.5$ & $3.5 \pm  0.6$  \\[0.1cm]
$\Phi_0\,\,[10^{-12} \text{cm}^{-2}\,\text{s}^{-1}\,\text{TeV}^{-1}]$  & $2.2 \pm 0.5$ & $2.4 \pm  0.7$  \\[0.1cm]
   R.A. $[^\circ]$ & $94.33 \pm 0.04$ & $94.38 \pm  0.07$ \\[0.1cm]
   Dec. $[^\circ]$ & $22.59 \pm 0.03$ & $22.58 \pm  0.05$ \\[0.1cm]
   $\sigma$ $[^\circ]$ & $0.26 \pm 0.03$ & $0.17 \pm  0.03$ \\[0.1cm]
\noalign{\smallskip}
\hline   
\hline 
\end{tabular}
\end{table}
To assess the morphology of the $\gamma$-ray emission, both Gaussian and Disk shaped spatial models were fit to the observed excess. As a spectral model, a simple power-law model of the form
\begin{equation}
\Phi(E) = \Phi_0 \left( \frac{E}{E_0} \right)^{-\Gamma}
\end{equation}
was adopted, with $E_0 = 1~\mathrm{TeV}$.

The parameters for both models can be seen in Table \ref{tab:fitparamters}. Although both models can describe the emission well, the disk model provided a marginally better fit with higher statistical significance. The significance of the Disk model is computed as $4.12\sigma$, compared to $3.97\sigma$ for the Gaussian model. 
After subtracting the emission from the data, a significance map with a correlation radius of $0.16^\circ$ was computed. From this map a 1D histogram of the residual significance in the region was constructed. The histogram was then fitted with a Gaussian model, and the mean and standard deviation of the residual data were derived. The residual significance in the region after the subtraction of the emission described by the Disk model is consistent with pure background fluctuations ($\mu \approx 0.02$, $\sigma \approx 0.89$), and no significant $\gamma$-ray emission remains. 

\begin{figure}
\centering
\includegraphics[width=0.5\textwidth]{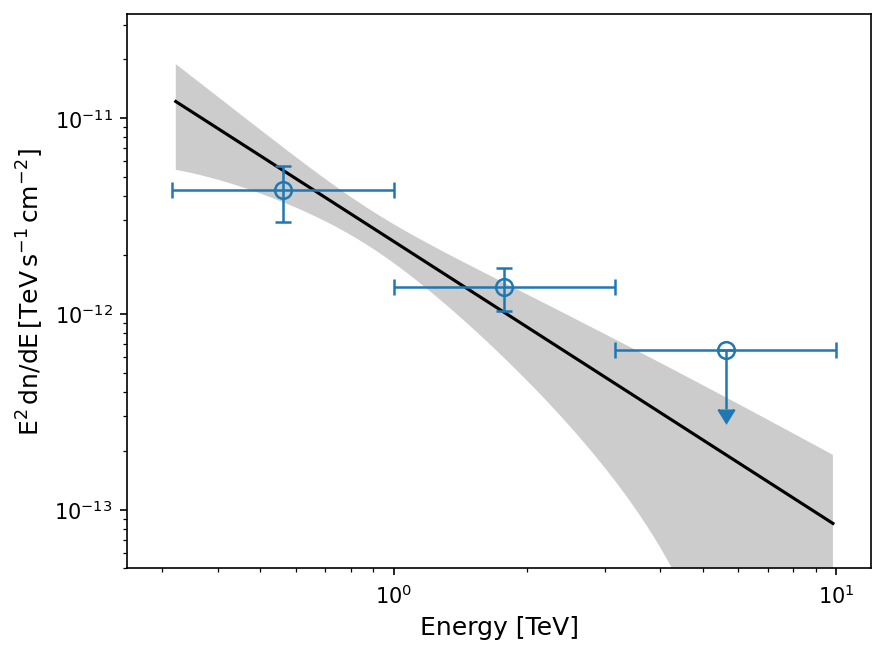}
\caption{SED of the emission detected from IC\,443, assuming a Disk morphology.}
\label{fig:SED}
\end{figure}

The values derived for both spatial models are in good agreement with previous measurements by Fermi-LAT, and VERITAS. They are also, within a $2\,\sigma$ uncertainty, compatible with the results derived by MAGIC \cite{MAGIC_2007}. The best-fit parameters for the preferred disk morphology model were $\Phi_0 = (2.2 \pm 0.5) \times 10^{-12}~\mathrm{TeV}^{-1}~\mathrm{cm}^{-2}~\mathrm{s}^{-1}$ and $\Gamma = 3.5 \pm 0.5$. The right panel of Figure \ref{fig:morphology} depicts a morphological comparison with previous analyses by Fermi-LAT, VERITAS, and MAGIC and includes both statistical and systematic errors for all source models. The H.E.S.S. systematics used in both panels of Figure \ref{fig:morphology} are from \cite{collaboration_resolving_2019}. The resulting spectral energy distribution (SED) for the power-law model is shown in Figure \ref{fig:SED}.
\section{Physical Modelling}
In order to derive the properties of the parent particle spectrum of the $\gamma$-ray emission from IC\,443, a hadronic origin via neutral pion decay, resulting from interactions of cosmic-ray protons with ambient gas in the remnant’s environment, was assumed. Given the well-established association of IC\,443 with molecular clouds and its composite morphology, this is strongly motivated both observationally and theoretically. 

\begin{figure}[h]
\centering
\includegraphics[width=0.65\textwidth]{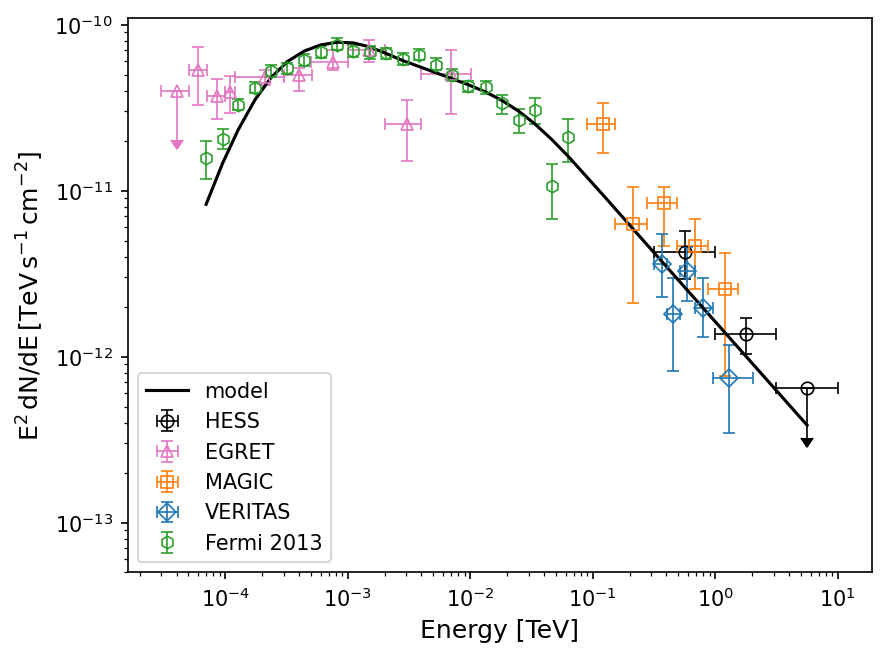}
\caption{SED derived in this work compared to the SED derived by EGRET \cite{esposito_egret_1996}, MAGIC \cite{MAGIC_2007}, VERITAS \cite{VERITAS_2009}, and Fermi-LAT \cite{Fermi_2013}. Additionally shown is the expected $\gamma$-ray spectrum produced by a proton population injected into the region. The injection parameters of this population have been derived through a fit to the SED. The SED derived by EGRET has not been used in the fitting process.}
\label{fig:SED_modeling}
\end{figure}

To this end, the SED shown in Figure \ref{fig:SED}, as well as data from Fermi-LAT, MAGIC, and VERITAS, were used to evolve a proton population and fit the resulting $\gamma$-ray spectrum to the observational data using version 0.10.0 of the \texttt{naima} Python package \cite{zabalza_naima_2015}. Due to the differing sizes of the regions used for spectral analysis, the SED for both MAGIC and VERITAS have been scaled to correspond to the flux expected for the source region used in this analysis. Two models were tested: a power law with exponential cutoff, and a broken power law. Both are typical parameterizations of cosmic-ray source spectra and reflect different underlying physical scenarios. The exponential cutoff model points towards a maximum energy due to acceleration limits, while the broken power law suggests a spectral steepening due to propagation effects or energy-dependent escape.

    We find the broken power law describes the data best with a reduced chi-squared value of $\nicefrac{\chi^2}{\text{dof}}=\nicefrac{81}{31}$ for the exponential cutoff powerlaw model and $\nicefrac{\chi^2}{\text{dof}}=\nicefrac{72}{31}$. The broken powerlaw function is of the form:
\begin{equation}
    f(E) = \left\{\begin{array}{ll}
    A\left(\nicefrac{E}{E_0}\right)^{-\alpha_1} & :E<E_{\mathrm{break}}\\
    A\left(\nicefrac{E_{\mathrm{break}}}{E_0}^{\alpha_2-\alpha_1}\right)\left(\nicefrac{E}{E_0}\right)^{-\alpha_2} & :E>E_{\mathrm{break}}
    \end{array}\right.
\end{equation}
The fitted proton spectral indices are $\alpha_1 = 2.32 \pm 0.02$ below the break and $\alpha_2 = 2.89 \pm 0.04$ above it, with a break energy at $E_{\mathrm{break}} = 168^{+57}_{-40}~\mathrm{GeV}$, a reference energy of $E_{\mathrm{0}} = 6.1^{+3.8}_{-2.3}~\mathrm{TeV}$ and a normalization factor of $\log_{10}(A) = 46.25\pm0.50$. These values are consistent with theoretical expectations for diffusive shock acceleration at SNR shocks followed by spectral softening due to energy-dependent diffusion or escape.


The derived proton population implies that IC\,443 is accelerating particles to at least several TeV. With a total energy in cosmic-ray protons (above $1\,$GeV) that is compatible with expectations from a typical supernova explosion ($\sim 10^{50}\,$erg), depending on the assumed gas density (taken to be $n_H = 20\,$cm$^{-3}$ based on CO and $21\,$cm observations \cite{Fermi_2013, Spectrophotometry_fesen}). This further strengthens the case for SNRs as significant contributors to the Galactic cosmic-ray population.

While the hadronic model was strongly favored in previous works, small contributions from leptonic processes cannot be excluded, particularly in localized regions near the associated pulsar wind nebula. Future high-statistics observations, especially in the transition region between Fermi-LAT and current IACT sensitivity, will help refine the spectral shape and disentangle emission components.
\section{Conclusion and Outlook}
In this work, we presented the first analysis of $\gamma$-ray emission from the supernova remnant IC\,443 in which morphology and spectral parameters were derived at the same time. For this analysis, archival H.E.S.S. data taken between 2004 and 2005, amounting to a total of 11.4 hours of livetime, was used. Although the limited data results in a detection significance of only $4.12\sigma$, this analysis yields results that are fully compatible with previously published measurements by Fermi-LAT, MAGIC, and VERITAS.


Spectral analysis yielded a steep power-law index of $\Gamma = 3.5 \pm 0.5$, and the broadband SED is best described by a broken power-law proton injection spectrum, consistent with pion-decay $\gamma$-ray production. The characteristic spectral features support the interpretation of hadronic origin, confirming previous evidence for IC\,443 as a site of cosmic-ray acceleration interacting with a dense molecular environment.

While this analysis does not provide an independent detection due to the limited exposure time, it demonstrates the consistency of H.E.S.S. observations with earlier results and validates the source morphology and spectrum using a fully independent data set and analysis pipeline. In this way, the study contributes to the reproducibility of gamma-ray astrophysical results, an increasingly important aspect in a field where complex instrument-specific analysis techniques can lead to subtle discrepancies. By confirming the presence and characteristics of IC\,443 emission across different instruments, observation periods, and analysis methods, this work supports a broader understanding of SNRs as sources of Galactic cosmic rays.

\bibliographystyle{JHEP}
\bibliography{references}

\providecommand{\href}[2]{#2}\begingroup\raggedright\begin{thebibliography}{10}

\bibitem{fesen_nature_1984}
R.A.~Fesen, \emph{The nature of the filaments northeast of the supernova
  remnant {IC} 443.}, \href{https://doi.org/10.1086/162142}{\emph{The
  Astrophysical Journal} {\bfseries 281} (1984) 658}.

\bibitem{torres_supernova_2003}
D.F.~Torres, G.E.~Romero, T.M.~Dame, J.A.~Combi and Y.M.~Butt, \emph{Supernova
  remnants and gamma-ray sources},
  \href{https://doi.org/10.1016/S0370-1573(03)00201-1}{\emph{Physics Reports}
  {\bfseries 382} (2003) 303}.

\bibitem{olbert_bow_2001}
C.M.~Olbert, C.R.~Clearfield, N.E.~Williams, J.W.~Keohane and D.A.~Frail,
  \emph{A {Bow} {Shock} {Nebula} around a {Compact} {X}-{Ray} {Source} in the
  {Supernova} {Remnant} {IC} 443},
  \href{https://doi.org/10.1086/321708}{\emph{The Astrophysical Journal}
  {\bfseries 554} (2001) L205}.

\bibitem{esposito_egret_1996}
J.A.~Esposito, S.D.~Hunter, G.~Kanbach and P.~Sreekumar, \emph{{EGRET}
  {Observations} of {Radio}-bright {Supernova} {Remnants}},
  \href{https://doi.org/10.1086/177104}{\emph{The Astrophysical Journal}
  {\bfseries 461} (1996) 820}.

\bibitem{Fermi_2010}
A.A.~Abdo, M.~Ackermann, M.~Ajello, L.~Baldini, J.~Ballet, G.~Barbiellini
  et~al., \emph{{OBSERVATION} {OF} {SUPERNOVA} {REMNANT} {IC} 443 {WITH} {THE}
  {FERMI} {LARGE} {AREA} {TELESCOPE}},
  \href{https://doi.org/10.1088/0004-637X/712/1/459}{\emph{The Astrophysical
  Journal} {\bfseries 712} (2010) 459}.

\bibitem{Fermi_2013}
T.F.-L.~collaboration, M.~Ackermann, M.~Ajello, A.~Allafort, L.~Baldini,
  J.~Ballet et~al., \emph{Detection of the {Characteristic} {Pion}-{Decay}
  {Signature} in {Supernova} {Remnants}},
  \href{https://doi.org/10.1126/science.1231160}{\emph{Science} {\bfseries 339}
  (2013) 807}.

\bibitem{MAGIC_2007}
M.~Collaboration, \emph{Discovery of {VHE} {Gamma} {Radiation} from {IC443}
  with the {MAGIC} {Telescope}},
  \href{https://doi.org/10.1086/520957}{\emph{The Astrophysical Journal}
  {\bfseries 664} (2007) L87}.

\bibitem{VERITAS_2009}
V.A.~Acciari, E.~Aliu, T.~Arlen, T.~Aune, M.~Bautista, M.~Beilicke et~al.,
  \emph{{OBSERVATION} {OF} {EXTENDED} {VERY} {HIGH} {ENERGY} {EMISSION} {FROM}
  {THE} {SUPERNOVA} {REMNANT} {IC} 443 {WITH} {VERITAS}},
  \href{https://doi.org/10.1088/0004-637X/698/2/L133}{\emph{The Astrophysical
  Journal} {\bfseries 698} (2009) L133}.

\bibitem{the_hess_collaboration_observations_2006}
T.H.~Collaboration and F.~Aharonian, \emph{Observations of the {Crab} {Nebula}
  with {H}.{E}.{S}.{S}},
  \href{https://doi.org/10.1051/0004-6361:20065351}{\emph{Astronomy \&
  Astrophysics} {\bfseries 457} (2006) 899}.

\bibitem{donath_gammapy_2023}
A.~Donath, R.~Terrier, Q.~Remy, A.~Sinha, C.~Nigro, F.~Pintore et~al.,
  \emph{Gammapy: {A} {Python} package for gamma-ray astronomy},
  \href{https://doi.org/10.1051/0004-6361/202346488}{\emph{Astronomy \&
  Astrophysics} {\bfseries 678} (2023) A157}.

\bibitem{gammapy_zenodo}
A.~Aguasca-Cabot, A.~Donath, K.~Feijen, L.~Gréaux, L.~Giunti, B.~Khélifi
  et~al., \emph{Gammapy: Python toolbox for gamma-ray astronomy},  June, 2023.
\newblock 10.5281/zenodo.8033275.

\bibitem{collaboration_resolving_2019}
H.E.S.S.~Collaboration, H.~Abdalla, F.~Aharonian, F.A.~Benkhali, E.O.~Angüner,
  M.~Arakawa et~al., \emph{Resolving the {Crab} pulsar wind nebula at
  teraelectronvolt energies},  Sept., 2019.
\newblock 10.48550/arXiv.1909.09494.

\bibitem{gamma-hadron}
S.~{Ohm}, C.~{van Eldik} and K.~{Egberts}, \emph{{{\ensuremath{\gamma}}/hadron
  separation in very-high-energy {\ensuremath{\gamma}}-ray astronomy using a
  multivariate analysis method}},
  \href{https://doi.org/10.1016/j.astropartphys.2009.04.001}{\emph{Astroparticle
  Physics} {\bfseries 31} (2009) 383}
  [\href{https://arxiv.org/abs/0904.1136}{{\ttfamily 0904.1136}}].

\bibitem{impact}
R.D.~{Parsons} and J.A.~{Hinton}, \emph{{A Monte Carlo template based analysis
  for air-Cherenkov arrays}},
  \href{https://doi.org/10.1016/j.astropartphys.2014.03.002}{\emph{Astroparticle
  Physics} {\bfseries 56} (2014) 26}
  [\href{https://arxiv.org/abs/1403.2993}{{\ttfamily 1403.2993}}].

\bibitem{model}
M.~{de Naurois} and L.~{Rolland}, \emph{{A high performance likelihood
  reconstruction of {\ensuremath{\gamma}}-rays for imaging atmospheric
  Cherenkov telescopes}},
  \href{https://doi.org/10.1016/j.astropartphys.2009.09.001}{\emph{Astroparticle
  Physics} {\bfseries 32} (2009) 231}
  [\href{https://arxiv.org/abs/0907.2610}{{\ttfamily 0907.2610}}].

\bibitem{de_naurois_high_2009}
M.~de~Naurois and L.~Rolland, \emph{A high performance likelihood
  reconstruction of gamma-rays for imaging atmospheric {Cherenkov} telescopes},
  \href{https://doi.org/10.1016/j.astropartphys.2009.09.001}{\emph{Astroparticle
  Physics} {\bfseries 32} (2009) 231}.

\bibitem{mohrmann_validation_2019}
L.~Mohrmann, A.~Specovius, D.~Tiziani, S.~Funk, D.~Malyshev, K.~Nakashima
  et~al., \emph{Validation of open-source science tools and background model
  construction in gamma-ray astronomy},
  \href{https://doi.org/10.1051/0004-6361/201936452}{\emph{Astronomy \&
  Astrophysics} {\bfseries 632} (2019) A72}.

\bibitem{li_analysis_1983}
T.-p.~Li and Y.-q.~Ma, \emph{{ANALYSIS} {METHODS} {FOR} {RESULTS} {IN}
  {GAMMA}-{RAY} {ASTRONOMY}}, {\emph{The Astrophysical Journal} {\bfseries 272}
  (1983) }.

\bibitem{zabalza_naima_2015}
V.~Zabalza, \emph{naima: a {Python} package for inference of relativistic
  particle energy distributions from observed nonthermal spectra},  Sept.,
  2015.

\bibitem{Spectrophotometry_fesen}
R.A.~{Fesen} and R.P.~Kirshner, \emph{Spectrophtometry of the supernova remnant
  {IC 443}}, \href{https://doi.org/10.1086/158534}{\emph{Astrophysical Jounral}
  {\bfseries 242} (1980) 1023}.

\end{thebibliography}\endgroup
\acknowledgments
Alison M. W. Mitchell, Lukas Großpietsch, and Tina Wach are supported by the Deutsche Forschungsgemeinschaft, DFG project number 452934793.
This contribution was co-funded by a program supporting faculty-specific gender equality targets at Friedrich-Alexander University Erlangen-Nürnberg (FAU).
The full acknowledgments of the H.E.S.S. Collaboration can be found at \url{https://hess.in2p3.fr/acknowledgements/}. 
\end{document}